# Spatial Variations in the Physico-chemical Variables and Macrobenthic Invertebrate Assemblage of a Tropical River in Nigeria.

\*Michael O. Omoigberale[1]; Ifeanyi M. Ezenwa[2]; Ekene Biose[3] and Ogaga Otobrise[3]

[1 & 4.] Hydrobiology and Fisheries Unit, Department of Animal and Environmental Biology, University of Benin, Benin City, Nigeria.

[2.] Department of Zoology and Environmental Biology, University of Nigeria, Nsukka, Enugu State, Nigeria

[3.] Department of Environmental Management and Toxicology, University of Benin, Benin City, Nigeria.

\*Corresponding author: +2348020755024; E-mail: omoigbe@uniben.edu

ORCID: 0000-0002-6491-0076

**Abstract**

An assessment of the spatial variation in the physico-chemical variables and macrobenthic invertebrates assemblage of a stretch of Ethiope River, Nigeria was investigated monthly for 6 months at five study stations. The results revealed that water temperature (26.10 - 31.20 $^0$C), pH (4.55 – 7.58), dissolved oxygen (5.20 – 9.30 $mgl^{-1}$) and bicarbonates (3.01 – 134.20 $mgl^{-1}$) were similar across the study stations. The highest value of conductivity (190 µS) and total dissolved solids (94.3 $mgl^{-1}$) which occurred as outliers were recorded at station 1. The physico-chemical variables showed no significant different ($P < 0.05$) across the study stations. The abundance of the macrobenthic invertebrates encountered comprises 61 taxa in 34 families. The relative abundance was highest at station 3 and the least value recorded at station 1 constituting 32.26% and 21.27% respectively of the total abundance of the macrobenthic invertebrates. The Diptera which accounted for 52.77% of the overall abundance was the dominant group encountered, while the Decapoda (10.97%) which was prevalent in station 1 and the Ephemeroptera (8.05%) which was common in stations 2, 3 and 4 constituted the subdominant group in this study. Over 68% of the oligochaetes recorded were restricted to station 3 except *Dero limnosa* and *Nais* sp that occurred in all the study stations. The overall abundance of the macrobenthic invertebrates was not significantly different ($p < 0.05$) among the study stations. Based on the canonical correspondence analyses (CCA), total dissolved solids (TDS), biochemical oxygen demand ($BOD_5$), bicarbonates and pH influenced the macrobenthic invertebrate assemblage. The study stretch of Ethiope River was reasonably free from gross pollution as indicated by the presence of Ephemeroptera, Tricoptera, Odonata and Decapoda at the study sites. However, there were indications of organic contamination as shown by the preponderance of naidid oligochaetes and certain dipteran species including the chironomids. This study revealed a high diversity of macrobenthic invertebrates, and this was attributed to the relative spatial homogeneity of the physico-chemical variables recorded at the study stations.

## Introduction

Freshwater resources which occupy a unique position in relation to other natural resources is the most ubiquitous substances on our planet: albeit in quantity and quality (Shiklomanov, 1998). Stream ecological studies have shown a strong correlation between environmental variations and macrobenthic invertebrate community composition (Faith and Norries, 1989). Biological communities including macrobenthic invertebrates are structured by synergy of environmental factors (Miserendino, 2001) and interactions amongst or between individuals and groups that make up populations, assemblages, communities and ensembles (Fauth et al., 1996). Macrobenthic invertebrates are diverse and respond to both natural and man-induced changes in the environment (Ndaruga et al. 2004).

The ability of macrobenthic invertebrates to give better understanding of changing aquatic environments than chemical and microbiological data, which at least give short-term fluctuations makes these unique organisms excellent bioindicators (Ravera, 2000; Ikomi *et al*., 2005). Other potential benefits of studies of macrobenthic invertebrates are the quick assessment of biological resources for conservation purposes and the detection of pollution through differences between predicted and actual faunal assemblages (Ormerod and Edwards, 1987). Several studies have implicated environmental parameters such as velocity, stream size, pH, conductivity, riparian forest, nutrients, amount of dissolved oxygen, and presence of impoundments in determining the structural assemblages of macrobenthic invertebrates (Ogbogu and Akinya 2001; Ogbeibu and Oribhabor 2002; Arimoro et al., 2011; Zabbey and Hart, 2014; Arimoro and Keke, 2017). Studies by Dimowo (2013) have shown that surface waters bodies such as streams, rivers and lakes among others are the most available sources of water used for domestic purposes in majority of rural communities in the developing countries especially in Sub-Saharan Africa. Regrettably, these water bodies serve as locations for the discharges of human sewage, refuse, waste waters from domestic activities, industries, abattoirs and consequently threatens the sustainability and functionality of these freshwater ecosystems (Adewoye, 2010).

Nigeria freshwater resources are contained within the extensive river systems, lakes, flood plains and reservoirs which constitute about 12.4% of the country surface area (Olaosebikan and Aminu, 1998). The river Ethiope where this study was undertaken is one of them with a total stretch of about 100 km in the Niger Delta region of Nigeria. The River is impacted by industrial discharges from its catchment, sewage, abattoir waste water, market municipal solid waste (MMSK), and storm runoff, while the inhabitants of the adjoining villages rely mainly on the river for their fishing activities, domestic water supply, sand mining and inter-village transportation (Omo-Irabor and Olobaniyi, 2007). Other studies in this river includes Adiotomre et al, (1999); Ikomi, (1996); Kaizer and Osakwe (2010), however, information on the relationship between environmental variables and macrobenthic invertebrates are far from adequate. This study evaluates the spatial variations in the physico-chemical variables in relation to the macrobenthic invertebrate communities in the freshwater reaches of Ethiope River, a tropical river in the Niger Delta region, Nigeria.

## Materials and Methods

### Study area and sampling sites

The study was conducted in the upper fresh water reaches of the river Ethiope, Nigeria. The river takes its source from a spring at Umuaja, from where it flows through the main towns of Umuaja, Umutu, Obiaruku, Abraka, Eku, Sapele and Koko in Delta State for over 100 km and discharges into the Atlantic Ocean via the Benin River (Fig. 1). The study area is located within the tropical equatorial region with two climatic regimes: the wet season (April – October), and the dry season (November – March). The average annual rainfall is about 2800 mm with temperatures varying between 23ºC and 33ºC in the afternoon and dropping to between 18ºC and 22ºC at night, with a mean annual relative humidity of between 50% and 75% (Omo-Irabor and Olobaniyi, 2007). The river substratum comprises predominantly fine sand mixed with mud and pebbles. Debris and decaying organic materials also constitute part of the substratum. Four sampling stations were selected for the study.

Station 1 which was designated as the reference station because it was less influenced by human activities was located about 50 m from the river source. The riparian vegetation at the station consists of *Bambusa vulgaris*, *Echinochloa pyramidalis and Elaeis guineensis*. The substratum consists predominantly of fine sand, mixture of mud and decaying organic materials at the bank with fallen leaves. Anthropogenic activities comprise mainly religious activities including traditional worship/sacrifices. Station 2 was located about 0.66 km downstream of station 1. The dominate human activities include bathing, laundry, washing of motorcycles, local sand dredging, and palm oil mill activities. The riparian vegetation consists of mixed forest vegetation and crop farming including *Bambusa vulgaris*, *Elaeis guineensis*, *Ananas comosus*, *Pteridium aquilinum*, *Manihot utillisima* and *M. esculenta*, *Carica papaya*, E*chinochloa pyramidalis*, *Cocos nucifera and Musa acuminate*. The aquatic vegetation consists mainly of *Vossia cuspidate, Nymphaea* sp., and *Pistia stratiotes*. The substratum consists of fine sand, coarse stones and decaying organic matter. Station 3 located 1.3 km downstream of station 2 is highly perturbed by numerous anthropogenic activities including processing of agricultural products, cassava mill, fishing, dumping of refuse, bathing, laundry, sand dredging and a channeled drainage system. The substratum composed of decaying plant materials mixed with mud, clay and fine sand at the mid channel of the river. The riparian vegetation consists of *Bambusa vulgaris*, *Ananas comosus*, *Pteridium aquilinum*, *Manihot utillisima, M. esculenta*, *Carica papaya*, *Echinochloa pyramidalis*, *Cocos nucifera*, *Saccharum* sinense and *Musa acuminate,* while the macrophytes consists primarily of *Nymphaea* sp. At station 4 sited at Umutu, about 2 km downstream of station 3. The riparian vegetation is mainly crop farming *Manihot utillisima* and *M. esculenta*. The substratum is predominantly organic matter mixed with mud, clay and fine sand while human activities comprises bathing, fishing, laundry, washing of vehicle and dumping of household waste.

**Water sampling**

Surface water samples for the determination of environmental variables were collected monthly over a 6 months duration between June and November, 2017. The following parameters; surface water temperature was measured using mercury-in-glass thermometer, while pH, electrical conductivity (EC) and total dissolved solids (TDS) were determined with Extech 500-meter probes (Exstik II). Dissolved oxygen, Biochemical oxygen demand ($BOD_5$) and bicarbonates ($HCO^-_3$) were determined according to the APHA (1998) methods. Analysis of all samples commenced within 24 hours of sampling.

**Macrobenthic invertebrates sampling**

Kick sampling method was adopted in the sampling of macrobenthic invertebrates. A D-frame net of 500 µm mesh within an approximately 25 m wadable portion of the river was deployed in the collection of macrobenthic invertebrates. A composite sample comprising four samples were collected at each station to represent a lone sample (Arimoro and Muller, 2010). As the substrate was agitated, the sampling moved gradually upstream and the samples collected were preserved in 10% formalin. In the laboratory, samples were washed in a 500 µm mesh sieve to remove sand, and the macrobenthic invertebrates were then sorted from the substrates, identified using relevant keys, and then counted.

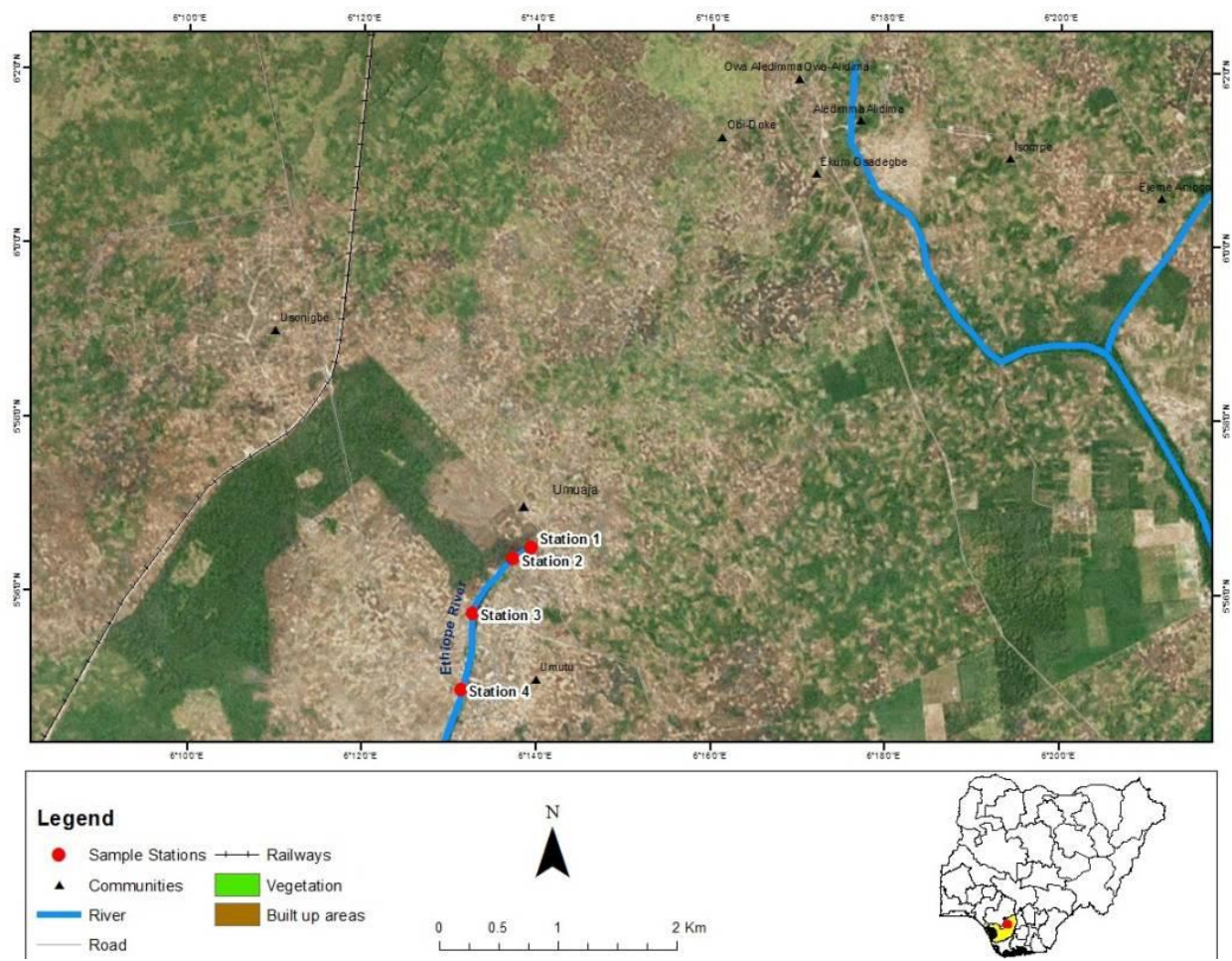

Figure 1: Study area showing sampling locations in the river Ethiope (Inset) Nigeria showing Delta State.

**Data Analyses**

The central tendency for each physico-chemical parameter was estimated for each station and an evaluation between stations using one-way analysis of variance (ANOVA). The variations in the physico-chemical parameters across the study stations were represented with boxplot and compared using one-way ANOVA (PAST version 3.14). Fixed effect ANOVAs were performed using dates as replicates and significant ANOVA ($P<0.05$). The correlation in the physico-chemical parameters in relation to the study stations were performed using non-metric multidimensional (NMDS) scaling techniques (PAST version 2.16).

Canonical correspondence analysis (CCA), a potent technique for simplifying complex data sets and, being a direct gradient analysis, it allows integrated analysis of both taxa and environmental data (ter Braak and Smilauer 2002), was used to evaluate relationships between macrobenthic invertebrate communities and the physico-chemical variables (PAST 3.14). prior to using the CCA and NMDS, species data were transformed by taking square roots to down-weight high abundances; the physico-chemical data were transformed by taking logarithms. Rare species (< 2% at sampling stations) were not included in the CCA determination. Species-environment correlation coefficients provided a measure of how well variation in community composition could be explained by individual physico-chemical variables. A Monte Carlo permutation test with 199 permutations was used to assess the significance of the canonical axes extracted.

Taxa richness (Margalef), diversity (Shannon-Wiener) Evenness and Dominance indices were assessed, and non-parametric analysis of variance (Kruskal Wallis) and Mann-Whitney pairwise comparison were adopted for inter-station comparison in the abundance of macrobenthic invertebrates across the study stations and to pinpoint station(s) of significant difference respectively (PAST version 3.14).

**Results**

The results of the central tendency including the range, mean and standard deviation of the physico-chemical variables at each study station are presented in Table 1. The values of temperature, pH, electrical conductivity, TDS, DO, $BOD_5$ and bicarbonate were not significantly different ($p > 0.05$) across the four study stations.

**Table 1: Summary of Physico-chemical variables at the study stations of Ethiope River, Nigeria. (June - November 2017)**

| Parameters (Unit) | Station 1 x̄±SD(Min-Max) | Station 2 x̄±SD(Min-Max) | Station 3 x̄±SD(Min-Max) | Station 4 x̄±SD(Min-Max) | p –Value | Permissible limit (SON*) |
|---|---|---|---|---|---|---|
| Temperature ($^O$C) | 28.58±1.78(26.10-31.00) | 28.71±1.88(26.10-31.30) | 28.76±1.76(26.40-31.20) | 28.71±1.65(26.40-31.10) | P > 0.05 | 35.00 |
| pH | 6.40±0.75(5.10-7.58) | 5.82±0.85(4.55-7.08) | 5.85±0.91(4.57-7.01) | 5.93±0.92(4.69-7.04) | P > 0.05 | 6.50-8.50 |
| Total dissolved solids (mg/l) | 30.29±26.72 (9.80-94.30) | 25.16±14.59 (9.60-51.50) | 26.81±17.73 (8.50-52.60) | 26.56±17.64 (8.80-52.700) | p> 0.05 | 500.00 |
| Conductivity (µs/cm) | 61.62±54.35(19.26-190.70) | 51.53±29.03(18.59-103.60) | 53.48±36.64(11.42-105.40) | 47.51±29.17(18.49-102.20) | P > 0.05 | N/A |
| Dissolved Oxygen (mg/l) | 6.62±0.98(5.30-8.40) | 6.37±1.26(5.20-9.30) | 6.39±0.75(5.30-8.00) | 6.77±0.90(5.50-8.70) | P > 0.05 | 7.50 |
| Biochemical oxygen demand$_5$ (mg/l) | 3.58±1.69(1.10-6.40) | 3.20±1.45(1.20-5.80) | 3.08±1.42(1.00-4.90) | 3.37±1.34(1.30-5.60) | P > 0.05 | 0.00 |
| Bicarbonate (mg/l) | 30.91±38.69(3.10-128.10) | 37.70±31.49(3.13-91.50) | 39.70±41.52(3.01-134.20) | 35.08±32.52(3.21-103.70) | P > 0.05 | 200.00 |

*Nigerian Standard for Drinking Water Quality. Standards Organization of Nigeria (SON), 2007

The spatial variations in the physico-chemical variables are depicted in box plots (Fig. 2). The water temperature showed wider range of variations as the lower and upper ranges across the stations were relatively equal to 26 ºC and 31 ºC respectively. The medians across the stations were relatively similar as higher levels of variations were observed above the $3^{rd}$ quartile than the $2^{nd}$ quartile (Fig. 2a). The spatial values of pH revealed relatively higher and closer concentrations at station 1 than at the other stations (Fig. 2b). At station 2, the values were equally distributed between the $2^{nd}$ and $3^{rd}$ quartiles, while higher variations in stations 3 and 4 were recorded in the $2^{nd}$ than in the $3^{rd}$ quartile.

The spatial variations of electrical conductivity and total dissolved solids (Fig. 2c and d) showed that the highest level of variation occurred at station 1, and these values occurred as outliers for both parameters. The distribution of values skewed towards the lower concentrations, while high variations were accommodated within the $3^{rd}$ quartile.

The median values showed that variations exhibited by dissolved oxygen (DO) concentrations are inverse of that of biochemical oxygen demand ($BOD_5$) (Fig. 2e). The minimum values of DO across the stations were $> 5$ mgl$^{-1}$ and a value $> 9$ mgl$^{-1}$ was obtained at station 2 as the peak concentration, however, this value was within the standard measure of dispersion and hence it was not detected as an outlier. More homogeneous values were obtained within the $2^{nd}$ and $3^{rd}$ quartiles of stations 3 and 4 than in stations 1 and 2. The pattern of $BOD_5$ variations maintained higher level of incoherence within the $2^{nd}$ and $3^{rd}$ quartiles than in the $1^{st}$ and $4^{th}$ quartiles across the study stations. At the upper limit, $BOD_5$ ($> 6$ mgl$^{-1}$, but $< 6.6$ mgl$^{-1}$) which occurred as the highest value was recorded at station 1, the lowest value was recorded at station 3 (Fig. 2f). Bicarbonate ($HCO_3^-$) are usually characterized with hydrogen ion concentration in the aquatic body. The values across the stations maintained high levels of variation above the $3^{rd}$ quartile, however, in stations 1 and 3, these elevated values which ranged within 125 – 135 mgl$^{-1}$ occurred as outliers. Generally, the concentrations of $HCO_3^-$ obtained across the stations were skewed towards low concentrations (Fig. 2g).

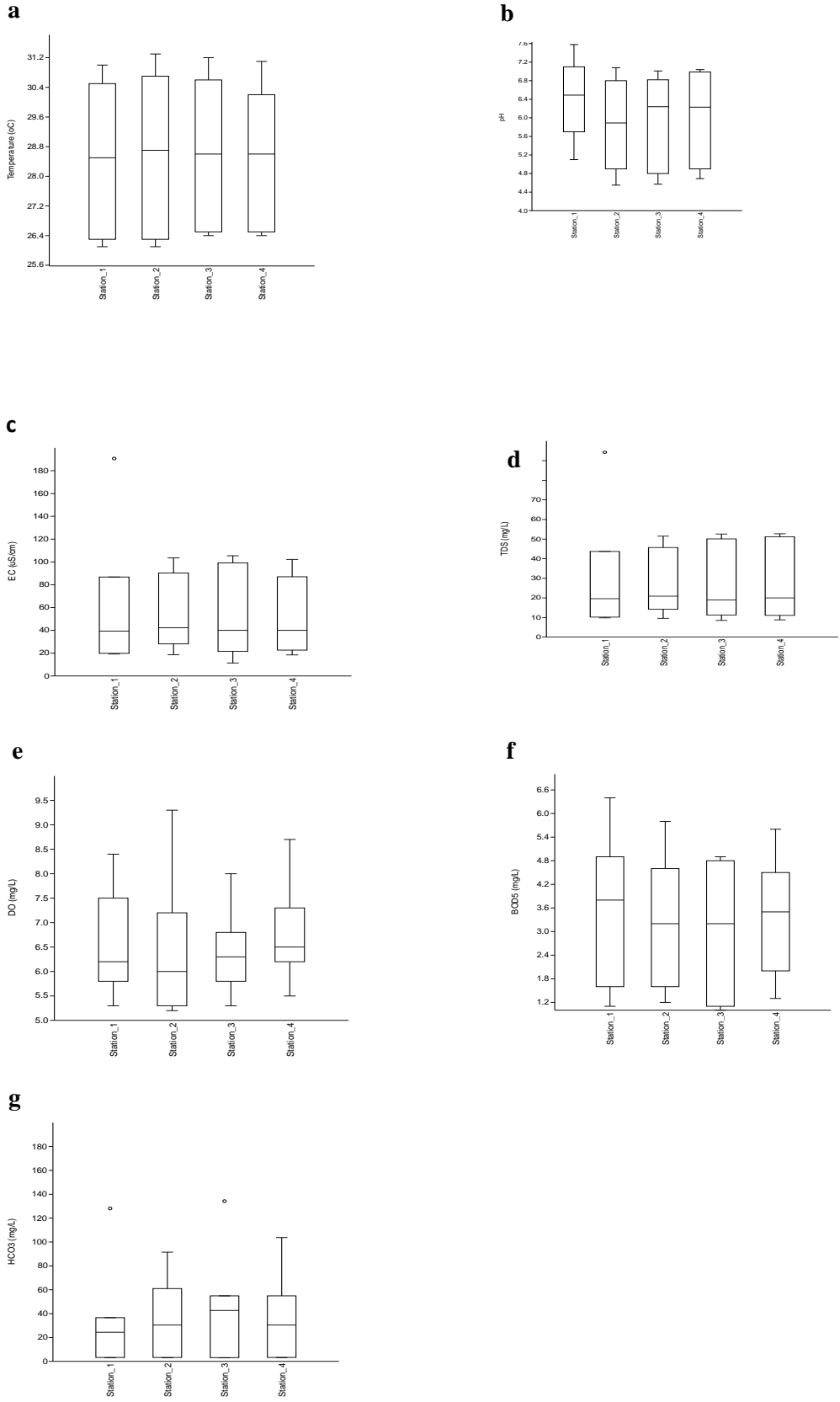

**Figure 2:** Box plot showing the spatial variation of the (a). temperature, (b). pH, (c) EC, (d) TDS, (e). DO, (f). $BOD_5$, (g) $HCO_3^-$ in Ethiope River from June to November, 2017

The similarity and correlations among the physico-chemical variables are shown using a non-metric multidimensional scaling (MDS) analysis (Fig. 3). The MDS ordination revealed 86% of the variance in the sample distance matrix (Coordinate 1 = 56%; Coordinate 2 = 30%, stress = 0.2147). The nucleated nature of the variables revealed high proximities among them. At 95% eclipses, all the physico-chemical variables were located within the various stations except pH and $BOD_5$ which were observed outside the spatial configuration of station 2. No significant similarity of these variables in station 2 with the levels in other stations was observed. Correlation of physico-chemical variables with stations location along the coordinate axes revealed changes in the gradients of the characterized variables. Coordinate 1 showed gradients in relation between temperature, total dissolved solid, electrical conductivity and biochemical oxygen demand, while Coordinate 2 revealed gradients associated with pH, dissolved oxygen, biochemical oxygen demand and bicarbonate. Temperature and pH were negatively correlated in coordinates 1 and 2 respectively.

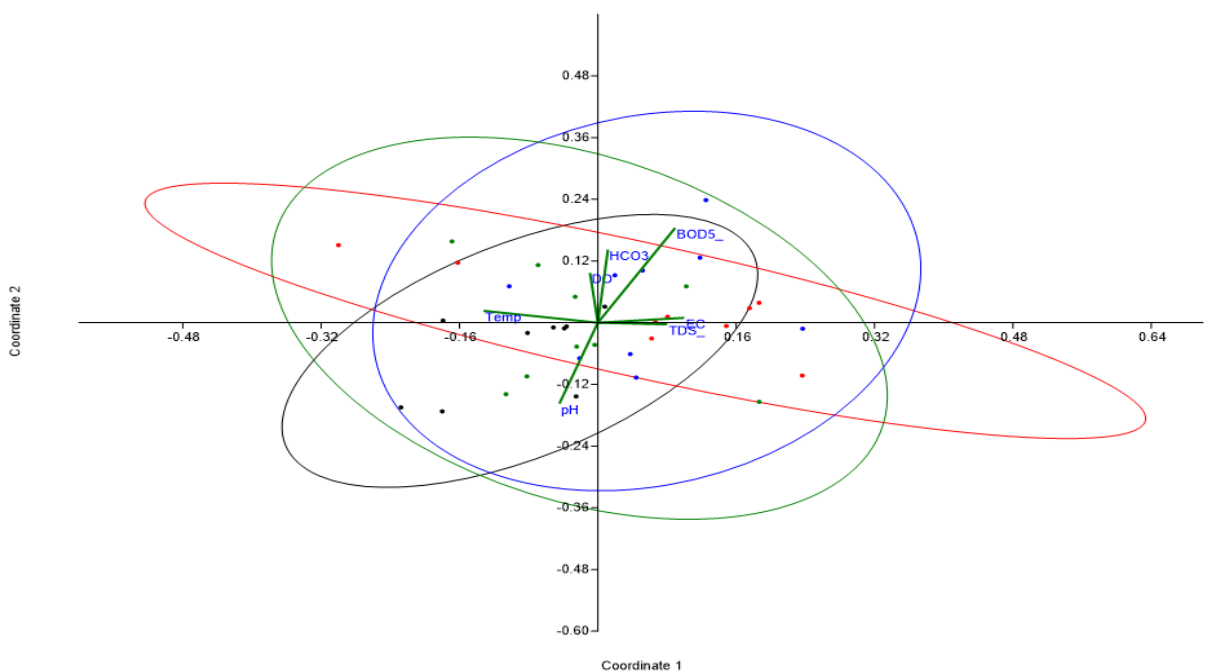

Figure 3: Two-dimensional pilot of non-metric Multi-Dimensional Scaling (NMDS) of the similarity and correlations among the environmental variables. Black dots and line (station 1); Red dots and line (station 2); Blue dots and line (station 3); Green dots and line (station 4).

## Macrobenthic invertebrate

A total of 61 taxa comprising 1,838 individuals belonging to 12 orders and 34 families were encountered during the study (Table 2). The relative abundance of the macrobenthic invertebrates revealed that station 3 had the highest value constituting 32.26% of the total abundance, while the least value (21.27%) was encountered at station 1. There was no significant difference (p >0.05) in the abundance of the macrobenthic invertebrates across the study stations, however, pairwise comparison using Mann-Whitney test revealed that the abundance at station 2 was significantly higher (p <0.05) than that of station 4. The Diptera which was the dominant group and commonest in all the study stations accounted for 52.77% of the overall abundance (Fig.4). It was represented by 13 taxa and the family chironomidae was the most abundant. Plesiopora represented by 7 species accounted for 11.83% of total abundance, while *Lumbricus* sp, *Enchytraeus* sp, *Aulophorous* sp and *Branchiodrilous* sp were recorded only in station 3, and *Pristina* sp encountered only at station 4. *Dero limnosa* which was the most abundant oligochaete was recorded in all the sampling stations. Decapoda and Ephemeroptera which comprises the subdominant groups accounted for 10.97% and 8.03% respectively of the total abundance. Higher abundance of Decapoda was recorded at station 1 in relation to stations 2, 3 and 4, while higher Ephemeroptera abundance were recorded at stations 3 and 4 and the least value was recorded at station 1. Rare taxa which were recorded in the orders - Prosopora, Hydrachnllae, Lepidoptera, Hymenoptera, Coleoptera, Tricoptera and Odonata, collectively accounted for 10.86% of the overall abundance recorded (Fig. 5).

Table 2: Abundance and distribution of macrobenthic invertebrates in Ethiope River, Nigeria. (June – November, 2017).

| Order | Family | Taxon | Station | | | |
|---|---|---|---|---|---|---|
| | | | 1 | 2 | 3 | 4 |
| **Oligochaeta** | Lumbricidae | *Eiseniella tetrahedra* | 3 | - | 20 | 8 |
| | Lumbriculidae | *Lumbriculus* sp | | | 4 | |
| | Enchytraeidae | *Enchytraeus* sp | | | 7 | |
| | Naididae | *Aulophorus* sp. | | | 4 | |
| | | *Branchiodrilus* sp | | | 7 | |
| | | *Dero limnosa* | 25 | 3 | 130 | 10 |
| | | *Nais* sp | 1 | 2 | 1 | 10 |
| | | *Pristina* sp | - | - | - | 1 |
| | | *Stylaria* sp | - | - | 2 | 15 |
| **Decapoda** | Atyidae | *Caridina gabonensis* | 49 | 3 | 39 | 21 |
| | | *Caridina africana* | 41 | - | 18 | 6 |
| | Desmocaridae | *Desmocaris trispinosa* | 14 | - | - | - |
| | Palaemonidae | *Euryrhynchina edingtonae* | 1 | 1 | 5 | - |
| | | *Leander tenuicornis* | 2 | - | 1 | - |
| | | *Macrobrachium* sp. | - | - | 1 | - |
| Hydrachnellae | Arrenuridae | *Arrenurus damkoehleri* | - | 11 | - | - |
| Araneida | Cybaeidae | *Argyroneta aquatica* | - | 2 | - | - |
| | Hygrobatidae | *Neumania* sp | - | 11 | - | - |
| Ephemeroptera | Baetidae | *Baetis tricaudatus* | - | 8 | 11 | 3 |
| | | *Centroptilum* sp | - | 3 | 36 | 40 |
| | | *Cloeon* sp | - | 7 | 10 | 20 |
| | | *Cloeon bellum* | - | - | 2 | 2 |
| | | *Cloeon cylindroculum* | 3 | 2 | - | - |
| | Ephemerellidae | *Ephemerella ignita* | 3 | - | - | - |
| Hemiptera | Gerridae | *Gerris* sp | - | 1 | - | - |
| | | *Rheumatobates* sp | 1 | 3 | - | - |
| | Mesoveliidae | *Mesovelia* sp | - | 6 | - | - |
| | Naucoridae | *Pelocaris femoratus* | 3 | 27 | 2 | - |
| | Veliidae | *Rhagovelia obese* | - | 59 | - | - |
| Lepidoptera | Pyralidae | *Nymphula* sp | - | 3 | 1 | - |
| Hymenoptera | Mymaridae | *Polynema natan* | - | 1 | - | - |
| Coleoptera | Chrysomelidae | *Donacia* sp | - | - | 1 | - |
| | Hydrophilinae | *Hydrophilus* sp | - | - | 1 | 1 |
| | Dytiscidae | *Hydroporus* sp | 4 | - | - | - |
| | | *Laccophilus variegatus* | 1 | - | - | - |
| | Elimidae | *Promeresia* sp | - | 1 | - | - |
| Trichoptera | Rhyacophilidae | *Rhyacophila fenestra* | 42 | 1 | - | 1 |
| Diptera | Ceratopogonidae | *Stilobezia antennali* | - | 3 | 1 | - |
| | Chironomidae | *Chironomus fractilobus* | - | - | 3 | 20 |
| | | *Chironomus transvaalensis* | 61 | 8 | 27 | 19 |
| | | *Cryptochironomus* sp. | 35 | 15 | - | 19 |
| | | *Polypedilum* sp. | 10 | 6 | 43 | 24 |

**Table 2 (cont.)**

| Order | Family | Taxon | Station | | | |
|---|---|---|---|---|---|---|
| | | | 1 | 2 | 3 | 4 |
| | | *Stictochironomus caffrarius* | 8 | - | 13 | 52 |
| | | *Tanytarsus* sp. | - | 8 | - | - |
| | Orthocladinae | *Corynoneura* sp. | 21 | 15 | 11 | 1 |
| | | *Cricotopus scottae* | 45 | 204 | 150 | 119 |
| | Tanypodinae | *Pentaneura* sp | - | - | 26 | - |
| | Culicidae | *Culex* sp. | - | - | 1 | - |
| | Simuliidae | *Simulium* sp | - | 1 | - | - |
| | Tanyderidae | *Protoplasa* sp | - | - | 1 | 2 |
| | Unidentified midges | | 1 | 5 | 5 | - |
| **Odonata** | Aeschnidae | *Aeschna* sp | 2 | - | - | - |
| | Gomphidae | *Phyllogomphus* sp | 4 | - | - | - |
| | Libellulidae | *Libellula* sp | 4 | 1 | - | - |
| | | *Orthetrum* sp | - | 3 | - | - |
| | | *Pachydiplax* sp. | 1 | - | - | - |
| | | *Sympetrum* sp | 2 | - | - | 2 |
| | Cordulegasteridae | *Cordulegaster* sp | 3 | - | - | - |
| | Macromiidae | *Macromia* sp | 1 | 11 | 2 | - |
| | Coenagrionidae | *Coenagrion tenellum* | - | 3 | 6 | - |
| | | *Coenagrion* sp | 2 | - | - | - |
| | | *Enallagma* sp | 1 | 7 | 2 | 8 |
| | | *Ischnura* sp | - | 1 | - | 3 |
| **Unidentified** | | | - | - | - | 2 |

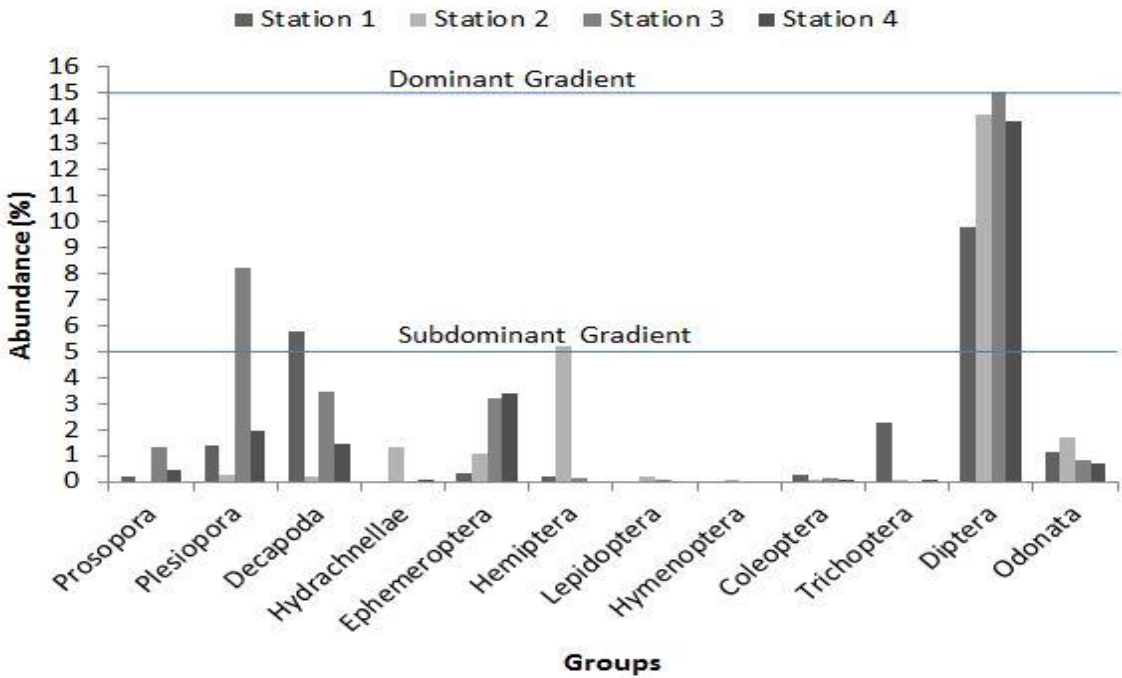

Figure 4: Overall variations in the abundance of macrobenthic invertebrate groups in Ethiope River, Nigeria (June – November, 2017).

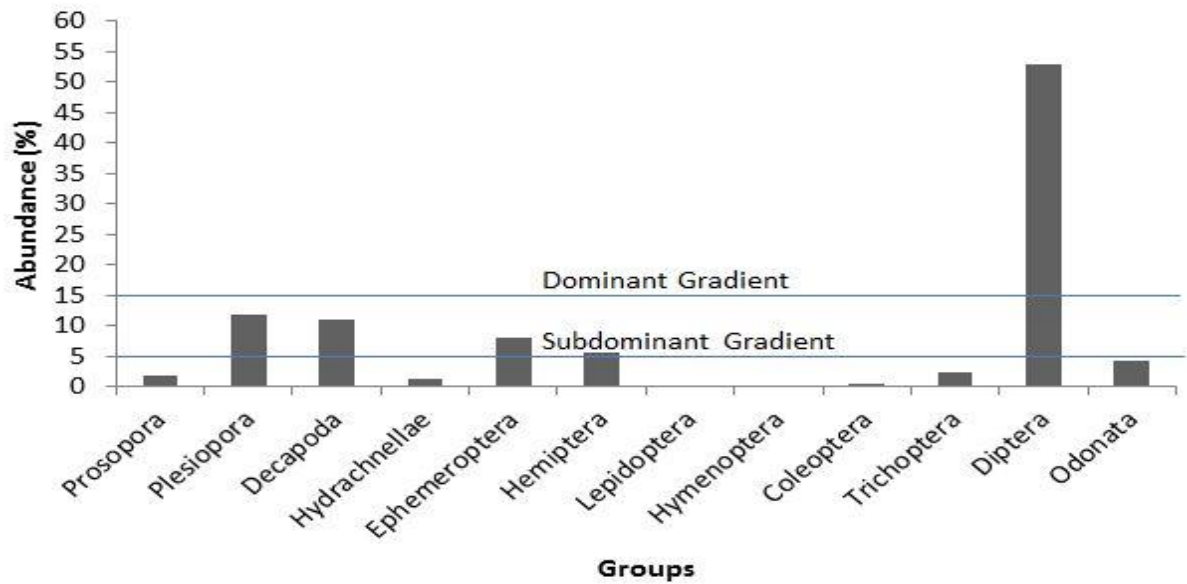

Fig. 5: Abundance and distribution of the orders of macrobenthic invertebrates in Ethiope River, Nigeria (June – November, 2017).

**Diversity Indices**

The values of the biological indices including abundance, number of taxa, Dominance, Shannon diversity, Evenness and Margalef estimated for the study stations are presented in Table 3. Dominance was highest at station 2 and lowest at station 1. Shannon diversity (H) recorded the least value (2.19) at station 2 and the highest value (2.65) at station 1. Margalef (richness) index (d) value was highest at station 2, followed closely by station 3. A much lower value of 3.83 was recorded at station 4. The evenness (E) values ranged from 0.26 to 0.50 at stations 2 and 4 respectively.

**Table 3** Summary of the diversity indices of macrobenthic invertebrates in Ethiope River, Nigeria (June – November, 2017).

| Indices | Station 1 | Station 2 | Station 3 | Station 4 |
|---|---|---|---|---|
| **Individuals** | 391 | 446 | 593 | 408 |
| **Taxa** | 30 | 37 | 35 | 25 |
| **Dominance_D** | 0.09 | 0.25 | 0.12 | 0.13 |
| **Shannon-Wiener H** | 2.65 | 2.19 | 2.51 | 2.48 |
| **Evenness_e^H/S** | 0.47 | 0.26 | 0.36 | 0.50 |
| **Margalef (d)** | 4.86 | 5.43 | 5.17 | 3.83 |

**Relationship between the Physico-chemical variables and macrobenthic invertebrates**

The CCA ordination showed relative relationships between taxa abundances and the physico-chemical variables (Fig. 6). The 2 canonical axes (axis 1: 40.19%; axis 2: 28.30%) accounted for 68.49 % of the variation in the weighted averages of 13 species and the seven physico-chemical variables, as well as the sum of all eigenvalues (0.693). Monte Carlo permutation test revealed that all the axes were significant. *Dero limnosa* and *Corynoneura* sp. which were restricted to stations 1, 3 and 4 were closely associated with $BOD_5$; $HCO^-_3$ and DO and negatively influenced by pH. *Cricotopus scottae*, *Centroptilum* sp., *Cloeon* sp., *Pelocaris femoratus*, *Rhagovelia obesa* were typical pointers of the environmental circumstances at stations 1, 3 and 4 including TDS and EC. *Stictochironomus caffrarius*, *Polypedilum* sp. and *Caridina gabonensis* which were positively influenced by pH did not show station preference, while *Corynoneura* sp , *Chironomus transvaalensis*,and *Caridina africana* which were positively influenced by water temperature were more restricted to station 1.

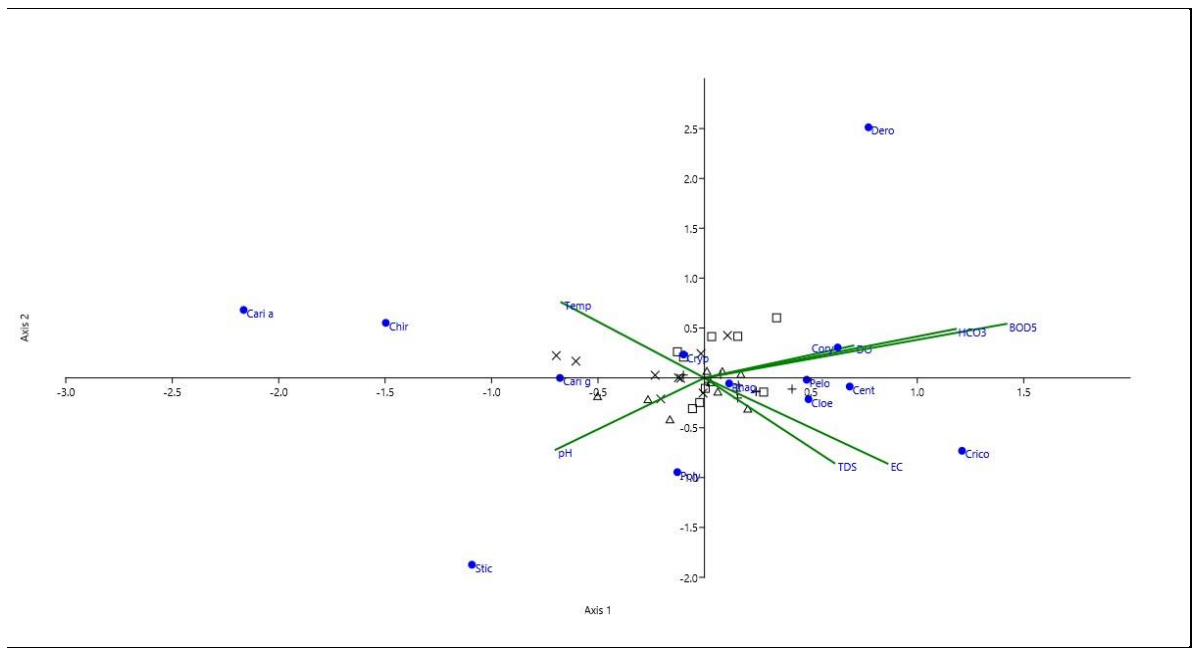

**Fig. 6:** Two-dimensional pilot of Canonical correspondence analysis (CCA) ordination. ×= Station 1; + = Station 2; □ = Station 3; Δ = Station 4; Chir = *Chironomus transvaalensis*; Cryp = *Cryptochironomus* sp.; Poly = *Polypedilum* sp.; Stic = *Stictochironomus caffrarius*; Crico = *Cricotopus scottae*; Cory = *Corynoneura* sp.; Dero = *Dero limnosa*; Cari g = *Caridina gabonensis*; Cari a = *Caridina africana*; Cent = *Centroptilum* sp.; Cloe = *Cloeon* sp.; Pelo = *Pelocaris femoratus*; Rhag = *Rhagovelia obesa*

## DISCUSSION

The Physico-chemical parameters of water bodies in addition to immediate substrate of occupation, natural geological features, land use pattern and other anthropogenic activities in the catchment influences the water quality and the community structure of macrobenthic invertebrates (Sundarmann et al, 2013; Arimoro and Keke 2017). The physical and chemical parameters of the Ethiope River revealed the surface water to be of relatively good quality as it was fairly homogenous across the study stations. The water temperature showed similar patterns across the study stations with a mean range of 28.58 °C to 28.76 °C. These values are typical of tropical Africa rivers (Ada et al., 2012; Andem et al. 2014). Temperature is a very important ecological factor whose variations can directly or indirectly affect some physico-chemical parameters including dissolved gases concentrations, pH as well as aquatic fauna which are predominantly ectothermic. Hydrogen ion concentration (pH) is one of the most common water quality parameters as it affects many chemical and biological processes in the water. The pH values recorded during the study is slightly acidic in nature, and this condition is typical of tropical forest streams. The acidic nature of Nigerian rivers had earlier been reported observed by Ajayi and Osibanjo, (1981).

Low concentration of electrical conductivity and elevated concentration of dissolved oxygen observed during the study reflects minimal impacts by anthropogenic activities at the study stretch. On the contrary, high concentrations of $BOD_5$ inferred that the study stations were moderately polluted by influx of organic matter from the numerous anthropogenic activities including washing detergents from laundry, processing of agricultural products, cassava mill and dumping of refuse into the river. Several reports have shown that anthropogenic activities ensuing from the release of organic contaminants into streams results in increase biological oxygen demand which subsequently affects the composition, distribution and abundance of macrobenthic invertebrates (Mwedzi et al. 2016)). Most rivers possess the ability to undergo self-purification at $BOD_5$ value of less than 4 $mgl^{-1}$, however this ability is compromised when it exceeds this value (Radojevic and Bashkin, 1999). Inhabitants of the catchment of rivers and streams depend on these water bodies for their water supply in many parts of Africa and other developing countries including Nigeria and these activities ultimately results in the contamination of these aquatic ecosystems (Arimoro and Muller 2010).

The macrobenthic invertebrates encountered in this study have been reported to be widely distributed in tropical Africa freshwater ecosystems including Nigeria inland water bodies (Omoigberale and Ogbeibu 2010; Ezekiel et al. 2011; Arimoro and Keke, 2017). A total of 61 macrobenthic invertebrate species were encountered in this study at Ethiope River. This number is similar to the 55 and 57 taxa recorded from Edo ecozone and the Osse River respectively (Olomukoro and Ezemonye 2007; Omoigberale and Ogbeibu, 2010), but much higher than the total macrobenthic invertebrate species from some rivers located in the southern part of Nigeria (Sikoki and Zabbey 2006). Although there were minimal differences in the macrobenthic invertebrates along the study stations, the highest abundance and diversity were recorded in stations 2 and

3. The high abundance and diversity observed at these stations could be ascribed to the heterogeneous nature of the vegetation of the littoral zone, which in turn served as an appropriate habitat for a more diverse macrobenthic invertebrate fauna.

The most widely distributed and abundant macrobenthic invertebrate fauna were the dipterans which were well represented all the study stations. The dominance of dipterans in many tropical assemblages of macrobenthic invertebrate groups have been documented (Ogbeibu & Egborge 1995). This is consistent with the findings of Ikomi et al. (2005) who reported the high abundance of Diptera in forest streams in Delta State, Nigeria. The occurrence of the Chronomidae in high abundance in this study is not unusual as this group of invertebrates have been reported to be common and are major components of tropical streams and related to the amount of organic matter from allochthonous input (Victor and Al-Mahrouqi, 1996). The Diptera, *Cricotopus* scottae was the most dominant species followed by *C. transvaalensis* and *Polypedilum* sp. They occurred in all the study stations in high numbers. This further confirms the leading position of chironomidae among dipterans, a common occurrence in both temperate and tropical waters (Lenat *et al*. 1981; Sharma *et al*. 1993). The dominance of the oligochaetes at station 3 including *Dero limnosa* and *Nais* sp. which were present in all the sampling stations are indicative of deteriorating biotic condition of the river especially at this station. The abundance of Oligochaetes has always been associated with muddy substratum rich in organic matter (Omoigberale and Ogbeibu, 2010). This explains why they were more abundant in station 3 and 4 where the substrata comprise mainly decomposing materials. Several studies have implicated increase density of oligochaetes in polluted rivers and streams Nigeria (omoigebrale and Ogbeibu, 2010). The presence and abundance of the Decapods, Ephemeroptera and Odonata; the less pollution-tolerant species in the study stations is an indication that these study sites are free from gross pollution. However, the absence of the ephemeropterans at station 1 is an evidence of organic pollution attributed to the input of organic matters resulting from cultural eutrophication which involves animal sacrifices during traditional worship frequently carried out at this station. These species are highly sensitive to drop in dissolved oxygen concentration and are not common in zones where oxygen levels are constantly low (Arimoro and Keke, 2017). The low species biomass and dominance in an water bodies could be an indication of an environmentally unfavourable condition, and according to Newall and Walsh (2005), a stressed biological population is characterized by reductions in diversity and population size.

The canonical correspondence analysis (CCA) ordination revealed that macrobenthic invertebrates were significantly associated with environmental variables determined in the Ethiope River. Electrical conductivity, biochemical oxygen demand ($BOD_5$) and pH were higher at station 1 than at the other stations, whereas temperature and bicarbonates ($HCO^-_3$) were lower at stations 1 and 4 than at the other stations. Station 1 was an outlier in the ordination studies, with a different macrobenthic invertebrate groups comprising many of the less-tolerant taxa such as *Rhyacophila fenestra* (Tricoptera),

*Caridina* sp (decapoda) and *Libellula* sp (odonata). The relative abundance of species of Tricoptera, Decapoda and Odonata are indicative of relatively good water conditions obtained at station 1. These groups have been used as indicators of clean water sites (Merritt and Cummins 1996). The dominance of pollution-tolerant dipteran groups such as *Cricotopus scottae*, *Chironomus transvaalensis* and *Polypedilum* sp, the oligochaetes including *Dero limnosa*, *Nais* sp, and *Eiseniella tetrahedra* which were relatively common across the study stations are early indications of pollution loads that can degrade the water quality and general ecological wellbeing of the river. Numerous investigations have documented increase in the density of these organisms in contaminated freshwater ecosystems in the southern part of Nigeria (Edokpayi, et al. 2004; Omoigberale, 2010;). Other unmeasured variables such as resource availability, biotic interactions and seasonality could be the reason while only 68.49% of the variation in the macrobenthic invertebrate groups was explained by the ordination, indicating that they could be important in determining the community structure of the macrobenthic invertebrate assemblage.

Diversity indices as estimated by taxa richness (d), general diversity (H') and Evenness (E) varied among the study stations. Taxa richness was significantly lower ($P > 0.05$) at station 4 than the other stations. This could be attributed to the water quality at this station which was relatively perturbed by human, agricultural and other activities at the river catchment. Shannon-Wiener diversity was not significantly different ($P > 0.05$) across the study stations and this could be a consequence of the ecological homogeneity across the stations. Evenness was highest at station 4 and lowest at station 2, and it was not significantly different ($P < 0.05$) in the study stations. Species diversity in streams have been reported to be highly variable and responsive to disturbances, availability of resource and the presence of appropriate environment (Fowler, 2002). Furthermore, equal or near equal opportunity of co-existence of many species is a consequence of high diversity. However, a decrease in diversity and corresponding increase in abundance of a limited number of species is a common response to environmental disturbance (Olomukoro & Ezemonye 2007).

The impact of human activities on freshwater bodies in the Niger Delta region of Nigeria has been on the increase and the Ethiope River is not an exception. This investigation provides information on the dynamics of the environmental variables which influences the community structure of the macrobenthic invertebrates in Ethiope River. The early warning signals of pollution loads a consequence of activities especially within the river catchment such as traditional worship/sacrifices, laundry, washing of motorcycles, local sand dredging and palm oil processing, cassava mill, dumping of refuse and fishing across the study stretch can potentially deteriorate the quality of the water and general ecological health of the river, and will require frequent monitoring to prevent further deterioration of the water quality.

**ACKNOWLEDGEMENT**

This research did not receive any specific grant from funding agencies in the public, commercial, or not -for-profit sectors.

This research did not receive any specific grant from funding agencies in the public, commercial, or not -for-profit sectors.